\documentclass{article}
\usepackage{PRIMEarxiv}
\usepackage[utf8]{inputenc}
\usepackage[T1]{fontenc}
\usepackage{hyperref}
\usepackage{url}
\usepackage{booktabs}
\usepackage{amsfonts}
\usepackage{nicefrac}
\usepackage{microtype}
\usepackage{fancyhdr}
\usepackage{graphicx}
\usepackage{CJKutf8}
\usepackage{float}
\usepackage{amsmath}

\title{Liver Cancer Knowledge Graph Construction based on dynamic entity replacement and masking strategies RoBERTa-BiLSTM-CRF model}

\author{
    Yichi Zhang\textsuperscript{1,a},
    HanLing Wang\textsuperscript{1,b},
    YongBin Gao\textsuperscript{1},
    XiaoJun Hu\textsuperscript{2,c},
    YingFang Fan\textsuperscript{2},
    ZhiJun Fang\textsuperscript{3}
    \\[2ex]
    \textsuperscript{1}Shanghai University of Engineering Science, China\\
    \textsuperscript{2}The Fifth Affiliated Hospital of Southern Medical University, China\\
    \textsuperscript{3}Donghua University, China\\[1ex]
    \textsuperscript{a}m320121358@sues.edu.cn,
    \textsuperscript{b}wanghailing@sues.edu.cn,
    \textsuperscript{c}xhxj2016@163.com
}

\begin{document}
\maketitle
\begin{CJK}{UTF8}{gbsn}
\begin{abstract}
Background: Liver cancer ranks as the fifth most common malignant tumor and the second most fatal in our country. Early diagnosis is crucial, necessitating that physicians identify liver cancer in patients at the earliest possible stage. However, the diagnostic process is complex and demanding. Physicians must analyze a broad spectrum of patient data, encompassing physical condition, symptoms, medical history, and results from various examinations and tests, recorded in both structured and unstructured medical formats. This results in a significant workload for healthcare professionals. In response, integrating knowledge graph technology to develop a liver cancer knowledge graph-assisted diagnosis and treatment system aligns with national efforts toward smart healthcare. Such a system promises to mitigate the challenges faced by physicians in diagnosing and treating liver cancer.

Methods: This paper addresses the major challenges in building a knowledge graph for hepatocellular carcinoma diagnosis, such as the discrepancy between public data sources and real electronic medical records, the effective integration of which remains a key issue. The knowledge graph construction process consists of six steps: conceptual layer design, data preprocessing, entity identification, entity normalization, knowledge fusion, and graph visualization. A novel Dynamic Entity Replacement and Masking Strategy (DERM) for named entity recognition is proposed.

Results: A knowledge graph for liver cancer was established, including 7 entity types such as disease, symptom, and constitution, containing 1495 entities. The recognition accuracy of the model was 93.23\%, the recall was 94.69\%, and the F1 score was 93.96\%.

\end{abstract}


\section{Introduction}
Cancer remains the foremost cause of mortality on a global scale. The 2020 Cancer Research Report indicates that liver cancer constitutes approximately 10.4\% of all cancer cases worldwide, amounting to nearly 10.1 million incidences \cite{ferlay2021cancer}. This rate positions it as the second most prevalent cancer type. Furthermore, liver cancer accounts for 6.3\% of all cancer-related fatalities, equivalent to approximately 5.5 million deaths, ranking it fifth in terms of cancer mortality rates.
\begin{figure}[htbp]
    \centering
    \includegraphics[width=1\textwidth]{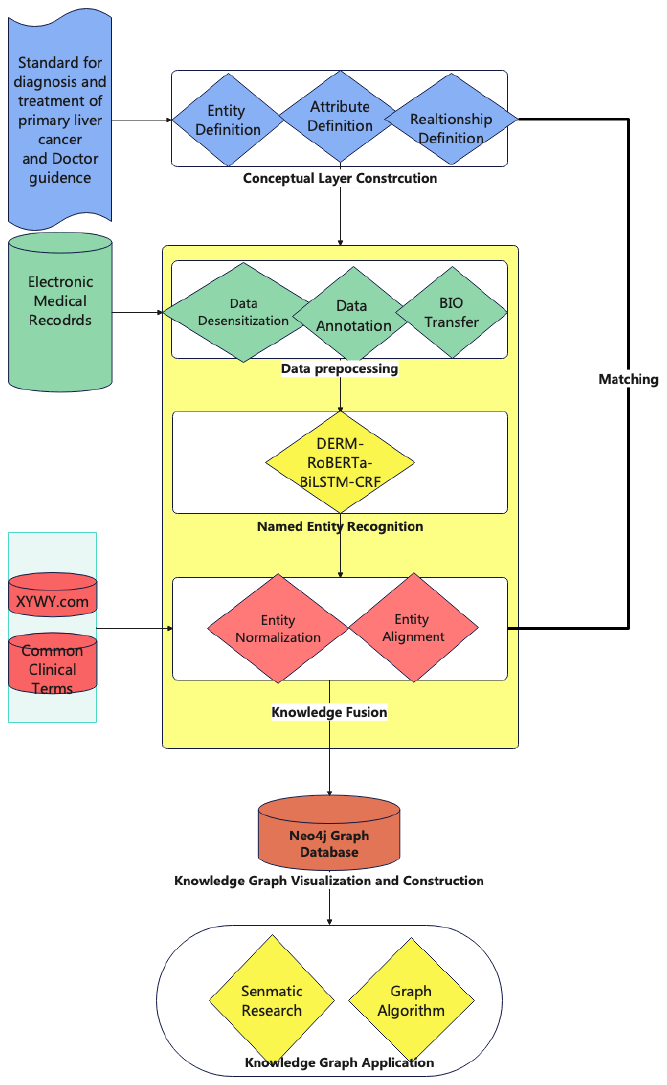}
    \caption[Short caption]{The proposed framework consists of 
    multiple steps including data preprocessing, entity recognition, and knowledge fusion.}
    \label{fig:framework}
\end{figure}
The Knowledge Graph (KG) was first proposed by Google \cite{10.3233/SW-190374} in 2012 as a structured knowledge representation of real-world entities and the relationships between them as graphical structures. Medical KG is a knowledge representation method that represents entities (e.g., diseases, drugs, symptoms, operation, etc.) and \ref{fig:framework}relationships between entities in the medical domain as structured graphs.

\section{Related Work}
\subsection{Medical Named Entity Recognition}
NER is an important branch in the field of natural language processing (NLP), which is mainly used to identify entities with specific meanings, such as diseases, symptoms, drugs, etc., from medical texts. In recent years, with the development of deep learning techniques, Chinese medical named entity recognition has made significant progress. Before deep learning techniques were widely used, Chinese medical named entity recognition mainly relied on rule-based and statistical-based approaches. Rule-based methods mainly use rules and dictionaries formulated by domain experts for entity recognition, such as regular expressions and dictionary matching. Statistical-based methods mainly include Hidden Markov Model (HMM)\cite{morwal2012named}, Maximum Entropy Markov Model (MEMM)\cite{saha2009feature}, and Conditional Random Field (CRF)\cite{sutton2012introduction}. Despite the advantages of simple implementation, higher accuracy, and lower computational resource requirements in some specific scenarios, these methods rely on rules and dictionaries formulated by domain experts, require a lot of human involvement, are difficult to deal with complex and flexible linguistic phenomena and have a weak generalization ability. With the development of word vector techniques (e.g., word2vec\cite{goldberg2014word2vec} and Glove\cite{pennington2014glove}), some breakthroughs have been made in the field of Chinese medical entity recognition (CMNER). Word vectors can characterize words into continuous high-dimensional vectors, thus improving the model's ability to capture the semantics of words. Through unsupervised learning, word vectors can be learned from a large amount of unlabeled data and the semantic relationships between words can be captured effectively. However, the word vector representation will be inaccurate for new words or words with multiple meanings. In recent years, deep learning techniques have been widely used in Chinese medical NER. The main methods include convolutional neural network (CNN)\cite{wu2017clinical}, recurrent neural network (RNN), and long short-term memory network (LSTM)\cite{sherstinsky2020fundamentals}. These methods can effectively capture the local features and long-distance dependencies of text to improve the accuracy of named entity recognition. Automatically learning local features and long-distance dependencies of text can provide better modeling of complex and flexible linguistic phenomena. However, these methods require a large amount of labeled data for training and the training process is time-consuming. With the emergence of pre-trained models, such as BERT\cite{devlin2018bert}, RoBERTa\cite{liu2019roberta}, and GPT3\cite{brown2020language}, Chinese medical NER research has entered a new era. These models are pre-trained on a large amount of unlabeled data such as Wikipedia and Reddit, which allows the word vectors obtained from pre-training to effectively capture rich semantic information. In the fine-tuning stage, the pre-trained models can be migrated to specific tasks, such as medical named entity recognition, which significantly improves the model performance.

\subsection{Construction of Medical Knowledge Graph}
Medical KG are characterized by dispersed knowledge distribution, distinctive syntax and non-standardized terminology, which makes the construction of medical KGs more difficult. In response to these challenges, researchers have undertaken diverse approaches to construct the Chinese medical KG. For instance, Zhang\cite{zhang2018generative} introduced a generative model, the Conditional Relationship Variational Autoencoder, aimed at reducing the workload in data preprocessing and manual annotation within the Chinese medical corpus\cite{zhao2020exploiting}. Deep learning models have been employed to enhance NER\cite{chowdhury2018multitask} and relation extraction in Clinical CEMRs[20]. Sheng [21] developed a comprehensive framework for a health knowledge graph, focusing on cardiovascular disease electronic medical records. Zhou[22] investigated the development and utilization of a 'knowledge-centric' traditional Chinese medicine knowledge graph, derived from ancient Chinese texts. However, one-way semantic relationships are inadequate to fully represent the complexities of patient medical processes. For instance, the semantic relationship between disease and examination encompasses not only investigating the disease but also revealing it through examination. To date, several medical knowledge graphs based on CEMRs have been developed, including those for conditions like hypertension and diabetes [23, 24]. Li[25] proposes a systematic approach to construct medical KG from large scale of EMRs. The constructed KG contains 9 entity types, totally 22,508 entities and 579,094 quadruplets. Xiu[26] proposes a framework for the construction of digestive system tumor knowledge graph based on CEMRs, and realized  construction of semantic-driven digestive system tumor knowledge graph(DSTKG).

To the medical knowledge graph application, exemplified by the semantic web for Chinese medicine, has garnered attention from both academic circles and the healthcare industry. Its utility in intelligent applications, such as analytical mining and drug recommendation, is particularly notable. For instance, Wang[27] propose a framework to conduct Safe Medicine Recommendation (SMR) and formulate it as a link prediction problem.

Our work on constructing liver cancer KG from CEMRs distinguishes itself from previous efforts in several key aspects: 1) It introduces the first knowledge graph specifically tailored for liver cancer, diverging from the general medical knowledge graphs typically seen in prior research; 2) involving normalizing and interconnecting entities like diseases, treatments, and surgeries in CEMRs with online medical knowledge bases; and 3) adding the downstream applications of the knowledge graph, rather than focusing only on the specific steps of construction KG as in previous work.

\section{Method}
In this section, we develop a systematic procedure to construct the liver KG from Chinese EMRs and the www.XYWY.com online resource. As illustrated in Figure 1, the procedure involves eight principal steps: 1) Conceptual Layer Design, 2) Data Preprocessing, 3) Entity Recognition, 4) Knowledge Fusion (KF), 5) Knowledge Dump, and 6) Knowledge Graph Application. It is important to note that steps 3 and 4 often demand extensive practical experience with Chinese EMRs and online resources
\subsection{Conceptual layer design}
In the medical domain, the conceptual layer design serves to abstract medical entities, organize and structure medical information systematically. This design aims to enhance the searchability and analyzability of healthcare data, thereby supporting medical decision-making, clinical research, and patient care. This study divides the conceptual layer into eight categories based on the professional medical website www.XYWY.com and clinicians' experiential knowledge.  The eight entity types and their specific entity content are shown in Table 1.
\begin{table}[H]
\centering
\caption{Eight categories of conceptual layer}
\begin{tabular}{|l|l|}
\hline
\textbf{Entity type} & \textbf{Entity content} \\
\hline
Patient & Patient ID and status (such as age$>$40) \\
\hline
Check & CT, MRI \\
\hline
Symptom & Left upper abdominal pain, vomiting \\
\hline
Diseases & Liver cancer \\
\hline
Condition & Smoking history \\
\hline
Operation & Cholecystectomy \\
\hline
Treatment & Laparoscopic right hepatic cancer resection \\
\hline
Body & Abdominal distension \\
\hline
\end{tabular}
\label{tab:conceptual-layer}
\end{table}
\subsection{Data preprocessing}
This study utilizes a dataset comprising real-world Chinese EMRs, content from the public professional website www.XYWY.com, and medical entity normalization documents such as CCMT to construct a liver cancer knowledge graph (KG).
 www.XYWY.com, a Chinese platform, provides comprehensive information on various diseases, including symptoms, diagnoses, treatments, medications, among others. From this source, we extracted semi-structured knowledge pertinent to liver cancer, highlighting symptom-disease and symptom-drug correlations. 
 \begin{table}[H]
    \centering
    \caption{Dynamic Entity Replacement and Masking Strategy}
    \label{tab:dynamic_entity_strategy}
    \begin{tabular}{|p{0.2\textwidth}|p{0.4\textwidth}|p{0.3\textwidth}|}
        \hline
        \textbf{Illustration} & \textbf{Sample} & \textbf{Probability} \\
        \hline
        Original text & 伴左上腹隐痛、呕吐、腹泻等 & \\
        \hline
        Dynamic Entity Replacement & 伴左上腹隐痛、呕吐、头痛等 & 30\%, random replace one entity \\
        \hline
        Dynamic Entity Masking & 伴左上[MASK]隐痛、呕吐、头痛等 & 30\%, if the entity's length $\leq$ 5, mask one word, otherwise mask 20\% word \\
        \hline
        Do nothing & 伴左上腹隐痛、呕吐、腹泻等 & 40\%, no change \\
        \hline
    \end{tabular}
\end{table}
\subsection{Data Preprocessing and Annotation}

\subsubsection{Data Source and Preprocessing}
The Electronic Medical Records (EMRs) used in this study were provided by Zhujiang Hospital of Southern Medical University in Guangzhou, China, covering the period from 2015 to 2020. These records contain comprehensive information about patient diseases, symptoms, and surgical procedures. To facilitate annotation and subsequent model training, we implemented the following preprocessing steps:

\begin{enumerate}
    \item Conversion of EMRs to readable and writable text formats
    \item Sentence reformatting based on punctuation
    \item Limitation of words per line to 50
\end{enumerate}

\subsubsection{Annotation Process}
For the annotation of Chinese EMRs, we utilized a client-side annotation tool, Colab. The annotation process focused on identifying and labeling several key entity types:

\begin{itemize}
    \item Diseases
    \item Symptoms
    \item Patient conditions
    \item Treatment plans
    \item Checkup information
    \item Surgery records
\end{itemize}

\begin{figure}[H]
    \centering
    \includegraphics[width=0.8\textwidth]{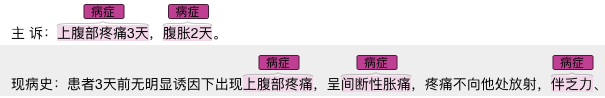}
    \caption{Example of disease entity annotation in Chinese EMRs}
    \label{fig:annotation-example}
\end{figure}

Figure \ref{fig:annotation-example} demonstrates an example of annotation for the ``disease'' (病症) entity category.

\subsubsection{Annotation Format}
\begin{figure}[H]
    \centering
    \includegraphics[width=0.8\textwidth]{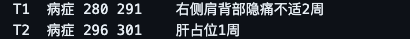}
    \caption{Example of Ann-Brat format annotation}
    \label{fig:ann-brat-format}
\end{figure}

The annotation results were saved in the Ann-Brat format, as illustrated in Figure \ref{fig:ann-brat-format}.

In this format:
\begin{itemize}
    \item ``T1'' indicates the first entity in the text
    \item ``disease'' (病症) specifies the annotation category
    \item The numbers 280 and 291 denote the start and end positions of the annotated entity
    \item The actual annotated text, ``右侧肩背部隐痛不适两周'' (dull pain in the right shoulder and back for two weeks), represents a specific disease entity
\end{itemize}

\subsubsection{Format Conversion}
To align with standard practices in named entity recognition tasks, we developed a Python script to convert the Ann-Brat format annotations to the widely-accepted BIO (Beginning, Inside, Outside) format.
\subsection{Named Entity Recognition}
Three types of NER methods are utilized to extract entities from EMRs: rule-based methods, machine learning-based methods, and deep learning-based methods. Rule-based methods prioritize the clinician's clinical experience, whereas machine learning and deep learning methods predominantly leverage data to train their models. Rule-based NER approaches, which depend on meticulously crafted semantic and syntactic rules. 
The BERT-BiLSTM-CRF model ranks among the most favored NER models in current use. In this study, the construction of DERM-RoBERTa-BiLSTM-CRF model is presented, replacing the pre-training model BERT with RoBERTa-wwm-large and adding DERM module to it. 
RoBERTa-wwm-large [30] is an open-source pre-trained model for Chinese dataset. Compared to BERT, RoBERTa-wwm-large removes the NSP (Next Sentence Prediction) task during the pre-training process, allowing the model to focus on a single sentence, reducing the noise in training and improving the model's performance on the NER task. In addition, the model uses the whole word masking strategy, which has a significant advantage in Chinese NER. This strategy involves masking all the Chinese characters forming a single word, thereby facilitating word-level information acquisition. As shown in Table 2, in the input text ‘with left upper abdominal pain, vomiting, diarrhoea, etc’(左上腹隐痛、呕吐、腹泻等). BERT may mask just one term of a character like ‘呕、泻’ thereby learning semantic representations at the level of character. In contrast, RoBERTa-wwm-large adopts a whole-terms masking strategy. It begins by segmenting the input text, then masks segments such as ‘left upper abdominal’(左上腹), ‘vomiting’(呕吐), and ‘diarrhoea’(腹泻). RoBERTa-wwm-large can learn term-level semantic representations, making it more effective for Chinese medical NER tasks. Consequently, this paper employs ROBERTa-wwm-large for vector representation of the input text.\begin{table}[H]
    \centering
    \caption{Pre-training mask strategy for BERT and ROBERTa-wwm-large}
    \label{tab:mask_strategy}
    \begin{tabular}{|p{0.2\textwidth}|p{0.7\textwidth}|}
        \hline
        \textbf{Illustration} & \textbf{Sample} \\
        \hline
        Original text & 伴左上腹隐痛、呕吐、腹泻等 \\
        \hline
        Segmented text & 伴 | 左 | 上腹 | 隐痛 | 、 | 呕吐 | 、 | 腹泻 | 等 \\
        \hline
        BERT-Base-Chinese masking strategy & 伴左上[MASK]隐痛、[MASK]吐、[MASK]泻等 \\
        \hline
        RoBERTa-wwm-ext-large's whole term masking strategy & 伴左[MASK][MASK]隐痛、[MASK][MASK]、[MASK][MASK]等 \\
        \hline
    \end{tabular}
\end{table}
DERM is a strategy used to process entities in natural language processing tasks. This strategy helps to improve the model's ability to recognize and process entities by dynamically replacing or masking entities in the input text. Firstly, a dictionary is constructed for each entity that has been manually annotated by categories such as diseases, symptoms, and treatment options, and then dynamic entity substitution and masking are performed on the text of the EMR during the training process, and the specific strategy is shown in Table 3. From Table 3, there is a 30\%possibility of using dynamic entity replacement and replacing one entity at a time. There is a 30\% possibility of using a dynamic masking strategy, i.e., masking several words in the entity words, and randomly masking one word in the entity if the length of the entity is less than or equal to 5, otherwise masking a 20\% number of words in the length of the entity.

Figure 5 illustrates the whole model framework, which comprises the DERM module, RoBERTa module, and BiLSTM-CRF module. ‘伴左上腹隐痛、呕吐、腹泻等’(with left upper abdominal pain, vomiting, diarrhea, etc.) is input Chinese sentence.  ‘B-症状’, ‘I-症状’and ‘O’ is entity recognition result.(‘B’ represents the start tag of entity, ‘I’ represents the inside tag of entity, ‘O’ represents the outside tag of entity and ‘症状’(symptoms) represents the entity types).

First, we use Roberta to perform automated segmentation of the input text and convert it into 1024-dimensional word vectors. Then, the word vectors are fed into a bi-directional long and short-term memory network (BILSTM). Conditional Random Fields (CRFs) are used as statistical learning models that can correct the final predicted irregular entity labels, thus improving the prediction accuracy of the model. The training, validation, and test sets are then divided in the ratio of 8:1:1 and placed into the deep learning model for training, validation, and testing

\subsection{Knowledge Fusion}

Knowledge Fusion (KF) is an integral component of Knowledge Graph (KG) construction. It encompasses the processes of extracting, integrating, correlating, and eliminating redundant information from multiple data sources to create a unified, comprehensive, and accurate knowledge representation. Through KF, we can effectively address several challenges in KG construction:

\begin{itemize}
    \item Data redundancy
    \item Information inconsistency
    \item Knowledge incompleteness
\end{itemize}

These improvements significantly enhance both the quality and usability of the resulting KG.During our entity extraction process, we observed significant variations in Electronic Medical Records (EMRs) for liver cancer patients. These variations stem from:

\begin{itemize}
    \item Differences in recording habits
    \item Diverse hospital operating guidelines
\end{itemize}

As a consequence, identical entities may be recorded using different terminology across various Chinese EMR reports, leading to inconsistent entity extraction.To address these challenges and facilitate better integration between EMRs and the www.XYWY.com knowledge base, entity alignment between the two knowledge bases becomes crucial. We employ the Term Frequency-Inverse Document Frequency (TF-IDF) method for this purpose.
TF-IDF is a statistical approach widely utilized in information retrieval and text mining. It evaluates the importance of a word in relation to a document within a corpus. This metric comprises two key components:

\paragraph{Term Frequency (TF)}
TF quantifies the occurrence rate of a word in a text, calculated as:

\begin{equation}
    TF(t,d) = \frac{t_d}{d_d}
\end{equation}

where:
\begin{itemize}
    \item $t_d$ represents the number of occurrences of word $t$ in document $d$
    \item $d_d$ represents the total word count in document $d$
\end{itemize}

The underlying principle is that the frequency of a word's appearance correlates with its importance in the text.

\paragraph{Inverse Document Frequency (IDF)}
IDF measures a word's distinctiveness across the entire corpus:

\begin{equation}
    IDF(t,D) = \log\frac{|D|}{|T|}
\end{equation}

where:
\begin{itemize}
    \item $|T|$ represents the number of documents containing word $t$
    \item $|D|$ represents the total number of documents in corpus $D$
\end{itemize}

IDF serves to reduce the weight of common words that appear across many documents but offer limited distinctiveness.

\paragraph{Combined TF-IDF Score}
The final TF-IDF value is obtained by multiplying TF and IDF:

\begin{equation}
    TF\text{-}IDF(t,d,D) = TF(t,d) \times IDF(t,D)
\end{equation}

This composite score effectively balances word frequency within documents against word distinctiveness across the corpus.

\section{Knowledge Graph Construction and Visualization}
The Neo4j Desktop version 1.5.8 (\url{https://neo4j.com/download/}) was used to construct the liver cancer KG in this study. In order to test the practicality and reasonableness of the constructed liver cancer KG, visual display and semantic query were conducted. Cypher was used for semantic query, and Match statements were used to query and filter irrelevant data. For instance, to query "what should you eat if you have liver cancer?", the Cypher statement would be:

\begin{verbatim}
MATCH (d:Disease {name: 'liver cancer'})-[:RECOMMENDED_FOOD]->(f:Food) 
RETURN f.name
\end{verbatim}

\section{Results}

\subsection{Entity Recognition}
Through the definitions in the conceptual layer of experts, we used Colaber's labeling tool to quantify each entity type, as shown in Table 1.

\begin{table}[h]
\centering
\caption{Statistics on the number of entities}
\begin{tabular}{lr}
\toprule
Entity Type & Number \\
\midrule
Patients & 304 \\
Physical Examination Information & 113 \\
Diseases & 449 \\
Patient status & 136 \\
Inspection results & 319 \\
Treatment & 171 \\
Surgical records & 3 \\
\bottomrule
\end{tabular}
\end{table}

The commonly used evaluation metrics for the named entity task in this study are accuracy P, recall R, and F1 score:A Python script was then used to convert the ann-brat format to the BIO format, which is often used as the standard format for named entity recognition tasks. The training, validation, and test sets were divided in the ratio of 8:1:1 and placed into the deep learning model for training, validation, and testing. The commonly used evaluation metrics for the named entity task in this study tour accuracy P, recall R, and Fi score. The accuracy rate indicates the proportion of true positive samples among the entities recognized by the model. The recall rate indicates the proportion of all positive samples that are correctly identified. The F1 score is the reconciled mean of the accuracy and recall rates. It is shown in the following equation.

\begin{equation}
P = \frac{\text{TP}}{\text{TP} + \text{FP}}
\end{equation}

\begin{equation}
R = \frac{\text{TP}}{\text{TP} + \text{FN}}
\end{equation}

\begin{equation}
F1 = \frac{2 \times P \times R}{P + R}
\end{equation}
\subsection{Comparison of Different Entity Recognition Methods}

We compared five modeling approaches: DERM-RoBERTa-Large-BiLSTM-CRF, Roberta-large-BiLSTM-CRF \cite{ref35}, DERM-BERT-Large-BiLSTM-CRF, and Word2vec-BiLSTM-CRF \cite{ref36}. The hyperparameters were set as follows:
\begin{itemize}
    \item Batch size: 40
    \item Epochs: 20
    \item Learning rate: $10^{-5}$
    \item LSTM hidden layers: 128
    \item Maximum sentence length: 50
\end{itemize}

\begin{figure}[h]
    \centering
    \includegraphics[width=\textwidth]{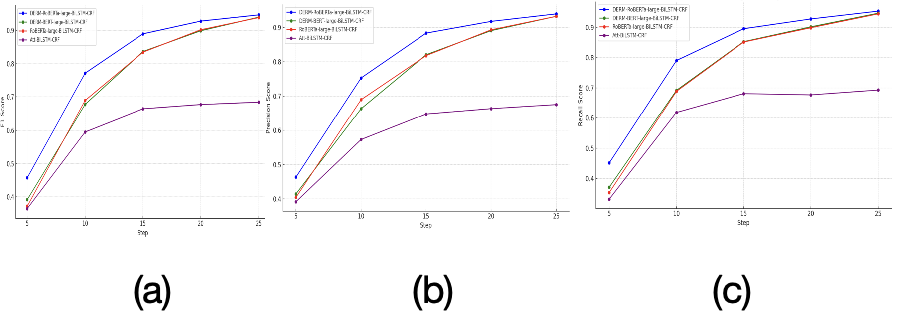}
    \caption{Entity recognition evaluation results of different models in validation set. (a) F1 scores, (b) Recall, and (c) Precision for different models.}
    \label{fig:model-comparison}
\end{figure}

The results show that RoBERTa-large-BiLSTM-CRF achieved the highest F1 score of 94.65\%, while the same model without DERM achieved 90.34\%, demonstrating a 4 percentage point improvement and proving the effectiveness of DERM.

\begin{table}[h]
    \centering
    \caption{Precision, Recall and F1 score in recognition of different entity types on DERM-RoBERTa-BiLSTM-CRF}
    \label{tab:entity-recognition}
    \begin{tabular}{lccc}
        \toprule
        \textbf{Entity Type} & \textbf{Precision} & \textbf{Recall} & \textbf{F1} \\
        \midrule
        Disease & 92.49\% & 93.02\% & 92.75\% \\
        Body Check & 91.59\% & 92.03\% & 91.80\% \\
        Symptom & 85.57\% & 86.56\% & 86.06\% \\
        Condition & 88.47\% & 88.98\% & 88.72\% \\
        Check & 92.13\% & 92.13\% & 92.13\% \\
        Treatment & 94.47\% & 93.86\% & 94.16\% \\
        Operation & 100\% & 100\% & 100\% \\
        \bottomrule
    \end{tabular}
\end{table}

Table \ref{tab:entity-recognition} illustrates the performance of DERM-RoBERTa-BiLSTM-CRF for seven significant entities within the test dataset. The Operation entity achieved the highest F1 score of 100\%, while the Symptoms entity recorded the lowest with an F1 score of 86.06\%. This result demonstrates the model's capacity for generalization in small sample datasets.
Here is the provided content converted into LaTeX format:

\subsection{Knowledge Fusion}

\begin{figure}[H]
    \centering
    \includegraphics[width=\textwidth]{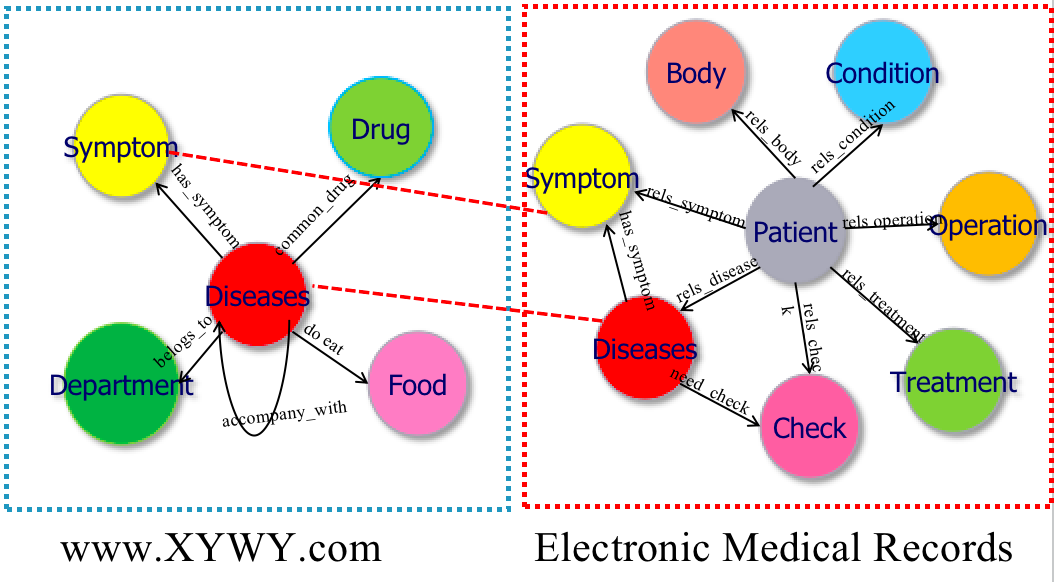}
    \caption{Knowledge graph fusion process}
    \label{fig:fusion_process}
\end{figure}

Figure~\ref{fig:fusion_process} illustrates the process of entity fusion in the knowledge graph of medicine-seeking. Initially, this knowledge graph is imported into Neo4j, consisting of six entity types: disease, food, department, drug, examination, and symptom. These entities are then normalized. The graph includes eight types of relationships: appropriate food (disease-food), contraindicated food (disease-food), department affiliation (department-belong), common drugs (disease-drug), diagnostic examinations (disease-examination), symptoms (disease-condition), complications (disease-disease), and departmental association (disease-department). 

For disease entity recognition, the EMR-based knowledge graph is first utilized to pinpoint all disease entities. Subsequently, a specific disease entity is queried using the www.XYWY.com disease entity library, employing a TF-IDF vectorizer to determine the highest cosine similarity match, with a minimum threshold of 0.8. For example, proto-hepatocellular carcinoma in the EMR aligns with primary hepatocellular carcinoma in www.XYWY.com. In Neo4j, the disease node of proto-hepatocellular carcinoma is replaced, establishing a new relationship with the patient ID and the corresponding entity node. Similarly, disease nodes in the original patient record are replaced with nodes from the medicine-seeking graph whenever possible, while unmatched nodes are temporarily retained for potential future modifications. 

Thus, TF-IDF integration of EMRs with the medicine-seeking knowledge graph enhances its completeness and forms a basis for subsequent tasks like knowledge graph analysis and reasoning.

\begin{figure}[H]
    \centering
    \includegraphics[width=0.8\textwidth]{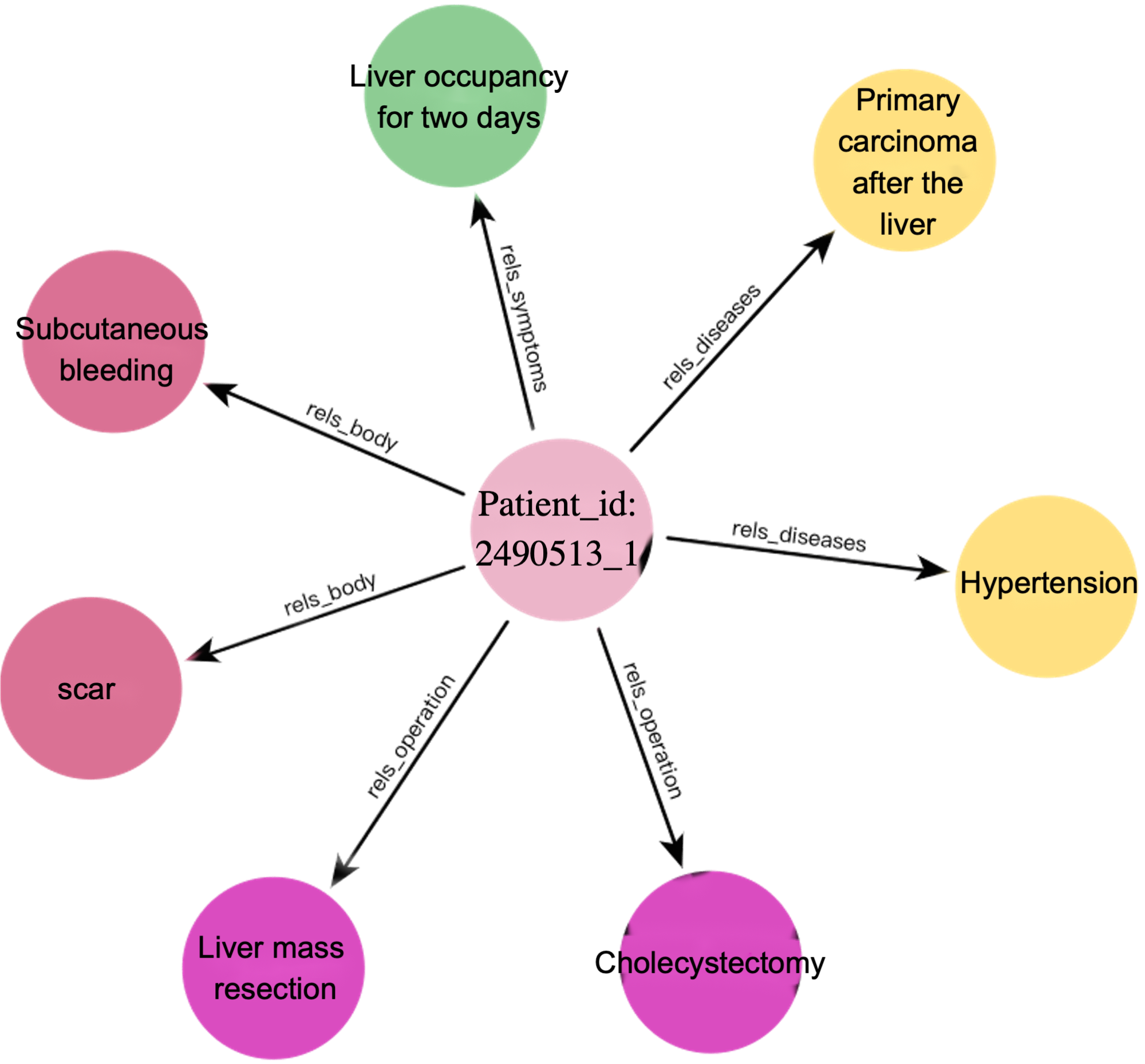}
    \caption{Knowledge Graph Prior to Knowledge Fusion for Patient ID '2490513\_1'}
    \label{fig:graph_prior}
\end{figure}

\begin{figure}[H]
    \centering
    \includegraphics[width=\textwidth]{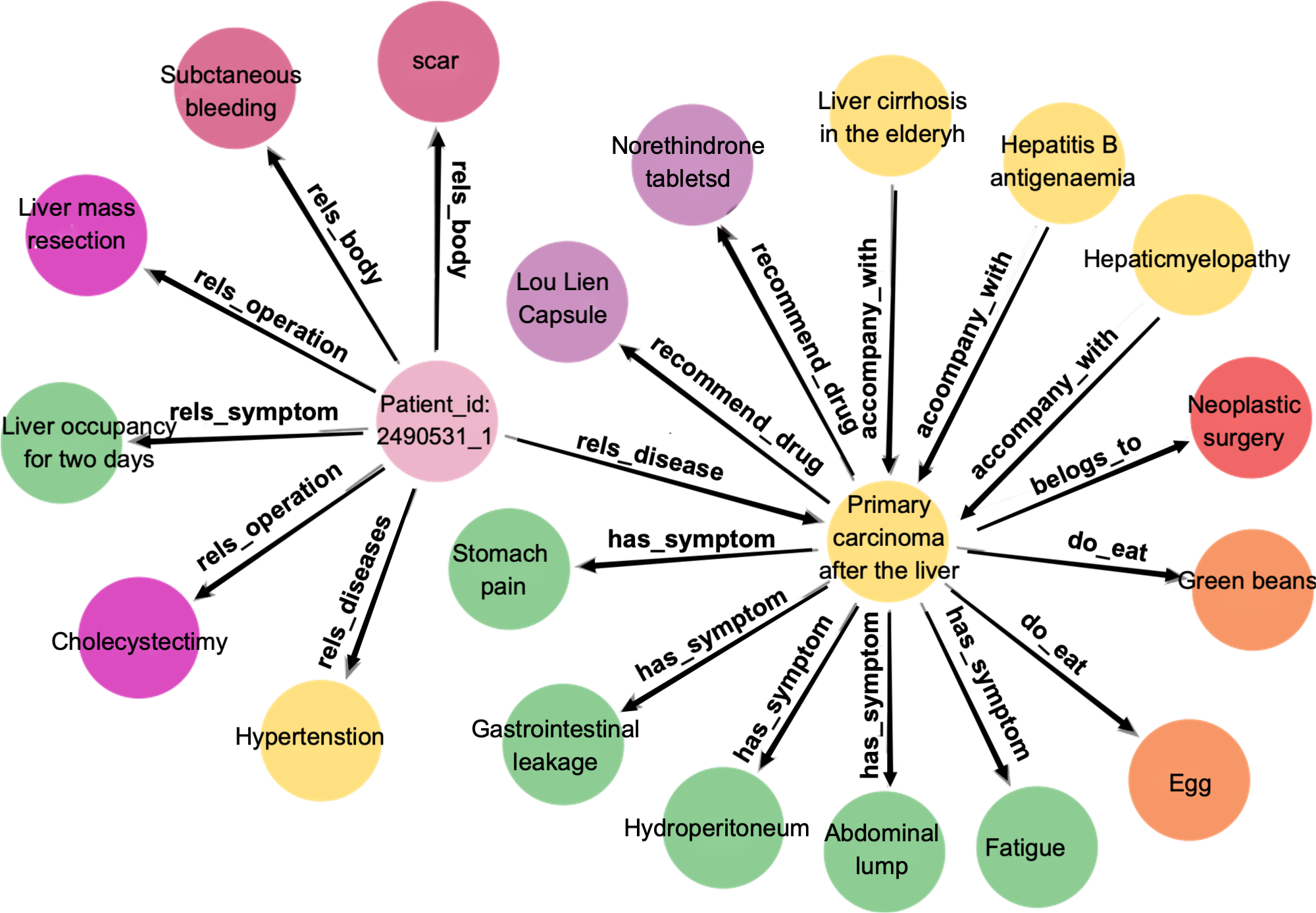}
    \caption{Knowledge Graph Posterior to Knowledge Fusion for Patient ID '2490513\_1' }
    \label{fig:graph_posterior}
\end{figure}

Figures~\ref{fig:graph_prior} and \ref{fig:graph_posterior} depict the Knowledge Graph Prior to Knowledge Fusion and Posterior to Knowledge Fusion for Patient ID '2490513\_1', respectively, presented in English and Chinese. The graphs show that entity matching significantly enhances the patient’s related entities and relationships, facilitating future downstream applications of the knowledge graph.

\section{Constructed Liver Cancer KG}

\begin{figure}[H]
    \centering
    \includegraphics[width=\textwidth]{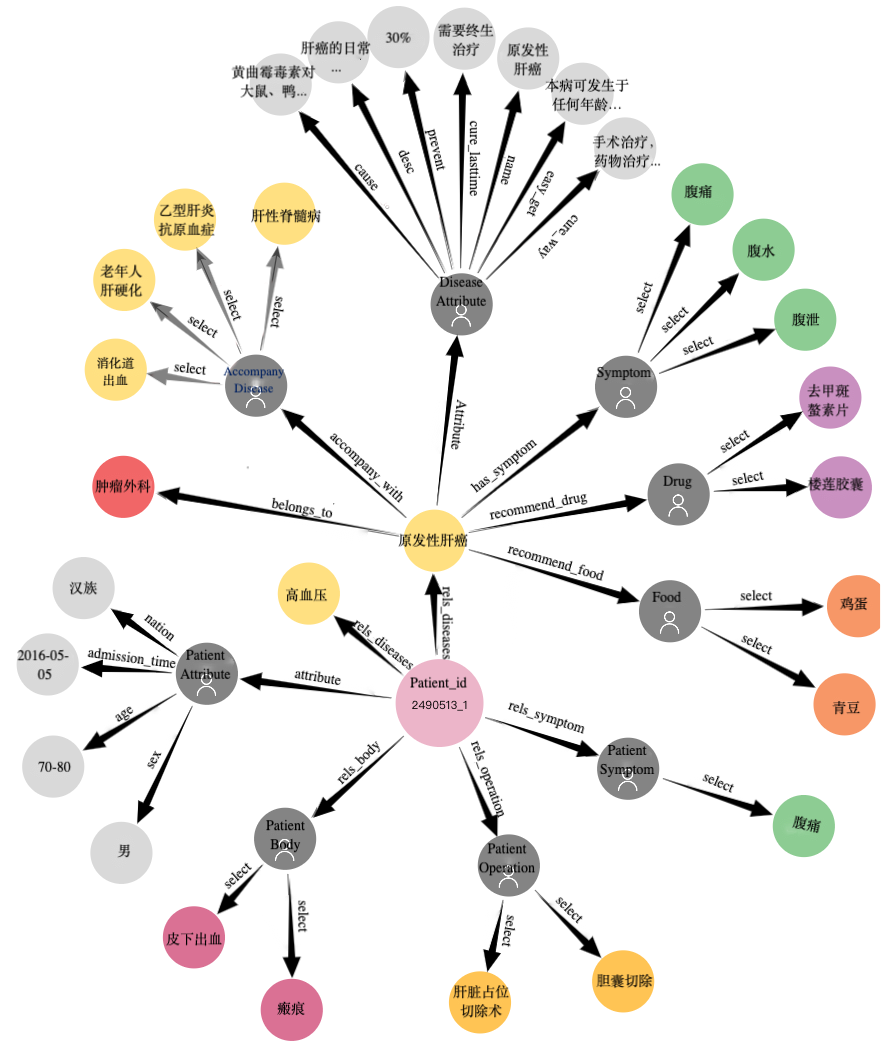}
    \caption{Overview of Liver Cancer Knowledge Graph}
    \label{fig:liver_cancer_kg}
\end{figure}

The constructed liver cancer KG contains 12 types of entities, totaling 46,365 entities and 296,655 triples. The triples cover relationships between patients and all 12 entity types. The number of entities and triples can be found in the table. In addition, the patient entity has basic attributes like nation, age, sex, and admission time. The disease attribute consists of the disease's name, description, prevention methods, cure time, treatment methods, and cause. Figure~\ref{fig:liver_cancer_kg} shows the attributes, diseases, and symptoms of patient No. 2490513\_3, and then the diseases are related to triples such as other complications.

\section{Application}

\subsection{Screening Basic Diseases for Liver Cancer Patients}
In clinical medicine, potential complication retrieval is essential. For doctors, identifying possible complications facilitates the development of more effective treatment plans, thereby mitigating the risk of complications. For patients, awareness of potential complications enhances their adherence to medical advice, encourages lifestyle modifications, and prompts timely reporting of any unusual symptoms. 

\begin{figure}[H]
    \centering
    \includegraphics[width=\textwidth]{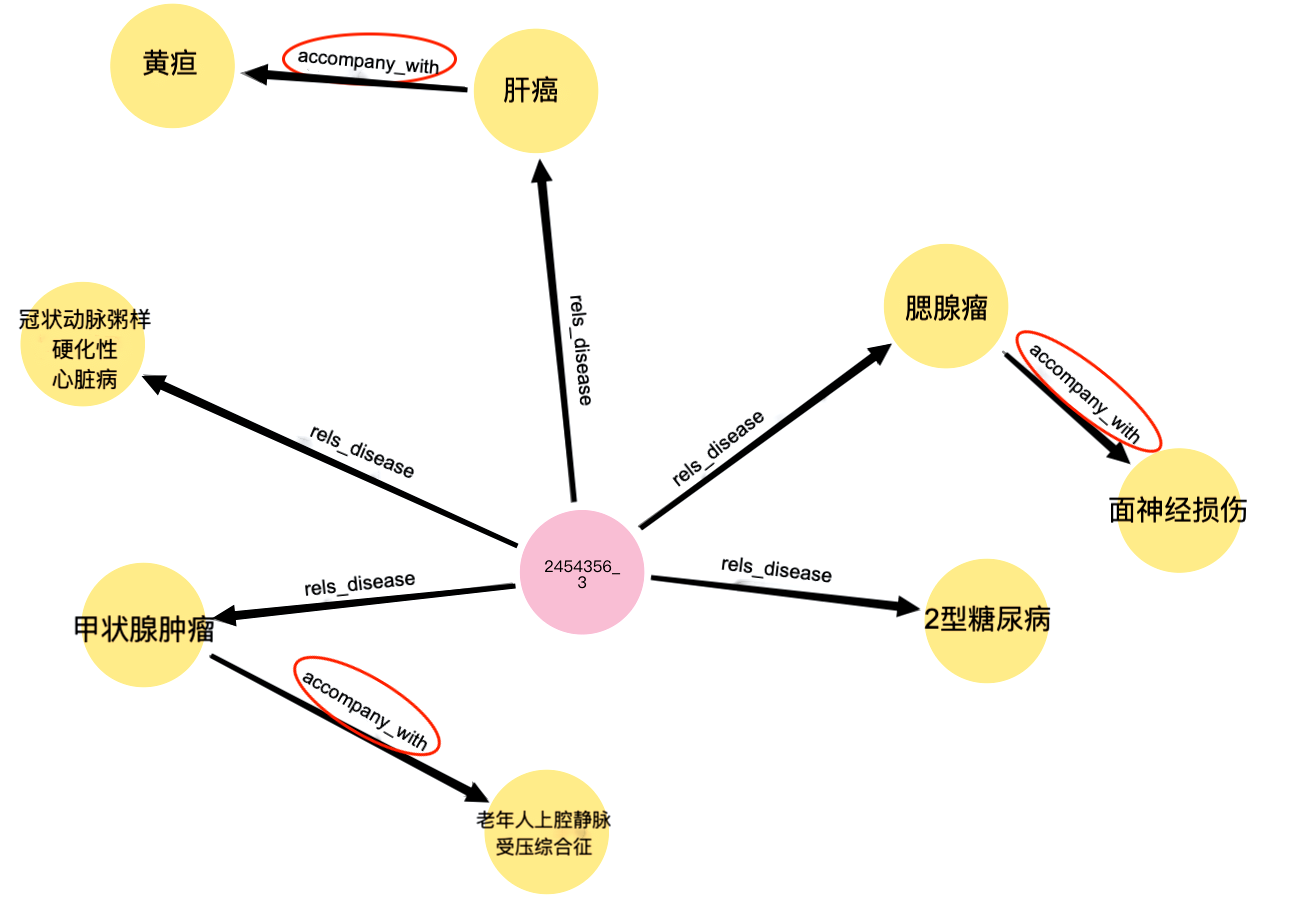}
    \caption{Semantic Research Result}
    \label{fig:semantic_search}
\end{figure}

Utilizing our liver cancer KG, we can efficiently associate patients with diseases and diseases with complications in the form of triples. This approach streamlines the retrieval of potential complications for subsequent patient assessments. Neo4j Bloom allows users to customize advanced query phases through the 'Search phrase' function. Users can pre-define static search templates without parameters or templates with parameters and then specify the phrases based on these templates to implement the query. As shown in Figure~\ref{fig:semantic_search}, it’s quick and easy to search for a disease that is accompanied by a patient's disease.

\section{Discussions}

\subsection{Performance Evaluation}
During the design process of the conceptual layer, experts in liver cancer diagnosis were invited to participate in the entity definition and structural design. In addition, the design of the conceptual layer was revised in multiple rounds with reference to domestic and international guidelines for liver cancer treatment. 

When the data layer was constructed, entity extraction was performed based on deep learning. To facilitate side-by-side comparison of model performance during model training, we set the same dataset of random seeds and parameters during cross-validation. The advantage of the DERM-RoBERTa-BiLSTM-CRF model is the use of RoBERTa-Large-wwm as our pre-training model, which is more suitable for the Chinese dataset because its masking strategy during pre-training is to mask according to the words instead of the form of individual words. Moreover, for a small number of noisy EMRs, the DERM module can effectively expand the dataset and reduce the noise.

Compared to the baseline Bert-Bilstm-Crf network model, our approach improved the F1 score by 4.3\% in both validation and test sets. However, this model requires the pre-formulation of DERM rules and the collection of standardized entities, resulting in a 20\% slower training time compared to the baseline model.

\subsection{Data Sources}
The dataset mainly consists of three parts: 
\begin{enumerate}
    \item 310 unstructured electronic medical records of liver cancer patients from ZhuJiang Hospital of Southern Medical University Hospital.
    \item A semi-structured medical dataset from www.XYWY.com, containing information about diseases, treatments, departments, and complications.
    \item Structured data from professional datasets such as CCMT, which provide professional clinical medical terminology.
\end{enumerate}
The first part contains rich information but also noisy data, requiring more work for entity extraction. The second part, from the XYWY.com website, requires crawler scripts for ternary information extraction. The third part provides highly professional data but requires a doctor’s guidance due to its complexity.

\section{Conclusion}

The primary contribution of this study lies in the development of a workflow for extracting knowledge graphs from Chinese EMRs to guide the construction and use of Traditional Chinese Medicine knowledge graphs for disease diagnosis and treatment. In this research, we designed the conceptual layer of the knowledge graph by referencing guidelines for primary liver cancer treatment and consulting with experts.

We employed the DERM-RoBERTa-BiLSTM-CRF model to extract entities such as patients, examinations, symptoms, and treatments from EMRs. Compared with the baseline model, the F1 score improved by 4.3\%. We standardized these entities using CCMT and integrated them with XYWY.com for knowledge fusion. Finally, the triples were imported into the Neo4j database. This study provides a reference for the rapid design and construction of knowledge graphs for the diagnosis and treatment of other diseases.

\bibliographystyle{unsrt}
\bibliography{templateArxiv}
\end{CJK}
\end{document}